\documentclass[twocolumn,showpacs,amsmath,amssymb]{revtex4}
\usepackage{dcolumn}
\usepackage{bm}
\usepackage{amsmath}
\usepackage{amsfonts}   
\usepackage{graphics}  
\usepackage{psfrag} 
\usepackage{graphicx} 
\usepackage{epsfig}    
 \usepackage{color}
 \usepackage{rotating}

\begin{document}

\title{On the distribution of career longevity and the evolution of home run prowess in professional baseball}
\author{Alexander Petersen,  Woo-Sung Jung, H. Eugene Stanley}

\affiliation{ Center for Polymer Studies and Department of Physics, Boston
University, Boston, Massachusetts 02215, USA}


\date{\today}

\begin{abstract}
Statistical analysis is a major aspect of baseball, from player
averages to historical benchmarks and records.  Much of baseball fanfare
is based around players exceeding the norm, some in a single game and
others over a long career.  Career statistics serve as a metric for
classifying players and establishing their historical legacy.  However,
the concept of records and benchmarks assumes that the level of
competition in baseball is stationary in time.  Here we show that
power-law probability density functions, a hallmark of many complex
systems that are driven by competition, govern career longevity in
baseball.  We also find similar power laws in the density functions of all
major performance metrics for pitchers and batters.  The use of
performance-enhancing drugs has a dark history, emerging as a problem
for both amateur and professional sports.  We find statistical evidence
consistent with performance-enhancing drugs in the analysis of home runs hit by players in
the last 25 years.  This is corroborated by the findings of the
Mitchell Report \cite{mitchell}, a two-year investigation into the use
of illegal steroids in major league baseball, which recently revealed that over 5
percent of major league baseball players tested positive for
performance-enhancing drugs in an anonymous 2003 survey.
\end{abstract}

\maketitle

Baseball is a game of legends, mystique, euphoria and heartbreak.  It is
also a game of numbers and records.  Here we analyze approximately 10,000 players
who ended their careers between the years 1920 and 2000, where 1920 is
the year widely considered as the beginning of the modern era of
baseball. We utilize Sean Lahman's
Baseball Archive \cite{Stat}, an exhaustive database consisting of Major
League Baseball player statistics dating back to 1871.  This database
was meticulously constructed, going so far as to extract data from old
newspaper reels.  We find that baseball players have universal properties
despite the distinct eras in which they played.  Specifically, we find
that the probability density functions of career totals obey scale-free power laws over a large range for
all metrics studied.  As usual, the probability density function $P(x)$ is defined such that the probability of observing an event in the interval $x+ \delta x$ is $P(x)\delta x$.
Power law density functions, $P(x)\sim x^{-\alpha}$, arise in many complex systems where competition drives the dynamics \cite{StanRev,Mantegna,Albert,sex,citations,wealth,Levy}.
A key feature of the scale-free power law is the
disparity between the most probable value and the mean value of the
distribution \cite{MEJN}.
For a Gaussian distribution, these two values coincide.
However, with a power law, the most probable value $x_{mp}= 1$, while the mean value
$\left<x\right> $ diverges for $\alpha \leq 2$.
Thus, in power law distributed phenomena, there are rare extreme events that are orders
of magnitude greater than the most common events.
This leads naturally to the notion of record events and the statistical
analysis of sample extremes \cite{Records}.
We begin this letter with an analysis of career longevity in American baseball.
Because the legacy of a player is based mainly upon his career totals, we also discuss the implications of the power-law behavior found in  common career metrics.
We conclude with empirical evidence, found in home run statistics, which is consistent with modern performance-enhancing factors including widespread use of performance-enhancing drugs.\\
\begin{figure}
\includegraphics[width=0.45\textwidth]{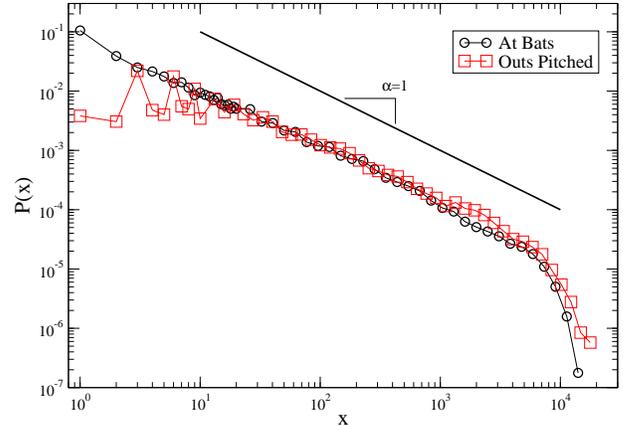}
  \caption{Probability density function  of player longevity. Longevity is
defined as the number of outs-pitched (pitchers), and the number of at-bats (all
batters) for players ending their career in the years 1920-2000.
  Power law extends over more than three orders of magnitude, with $\alpha \approx 1$.
  For reference: straight line represents the power-law $P(x) \sim x^{-1}$.
}
\label{long} 
\end{figure}
In Fig. \ref{long} we present the longevity of a player's career measured
in at-bats (AB) and innings-pitched measured in outs (IPO).  For these
two metrics, we find truncated power-law distributions that range over
three decades, marked by a sharp exponential cutoff at a value
corresponding to around twenty seasons.  It should be noted that unlike
a complete power-law distribution with $ \alpha \approx 1$, which has a
divergent first and second moment, a truncated distribution has a
definite mean and second moment.  To our surprise, we find that the
distributions for career longevity have their maxima around 1
appearance.  This implies that most players who make it to the major
leagues do not remain for very long, possibly making their professional
debut and exit in a single pinch-hit or relief appearance.  This leads
to a perplexing feature of scale-free power laws, namely that it is just
as hard to reach your 10th appearance from your debut appearance as a
rookie as it is to reach your 10,000th appearance from your 1000th
appearance as a seasoned veteran.  In other words, the ratio $
{P(x_{2})}/{P(x_{1})}= ({x_{2}}/{x_{1}})^{-\alpha}$ depends only on the
scale-free ratio of ${x_{2}}/{x_{1}}$ and the universal exponent $
\alpha$.  This raises a fundamental question addressing longevity in American baseball: How is it possible that the same level
of competition can eliminate some players after one appearance while
sustaining others for more than two decades?  American baseball has a 3-tier farm system, collectively known
as the minor leagues.  These developmental leagues filter talent up to
the major leagues, with only the best players staying at the major
league level.  Occasionally there are opportunities for minor league players
to be promoted to the major leagues for short unguaranteed stints,
either if their major league affiliate has a roster vacancy due to injury or if their
major league affiliate is not in a position to make the post-season.
The long regular season provides ample opportunity for these major league tryouts, thus accounting for the high frequency of short careers.\\
\begin{figure}[t]
\includegraphics[width=0.48\textwidth]{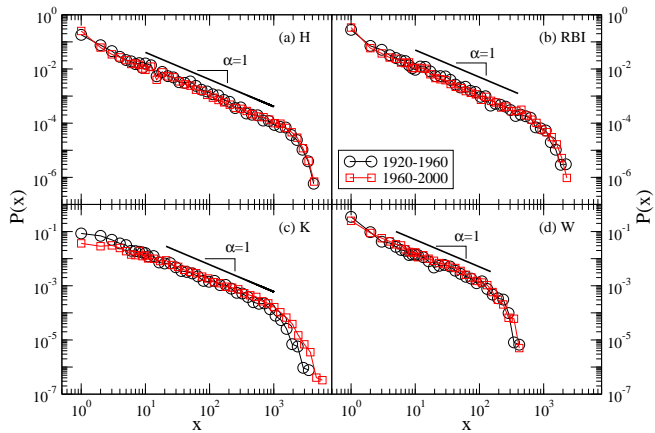}
  \caption{Probability density function of career statistics in four categories.
(a) Hits, (b) Runs Batted In, (c) K (strikeouts),
(d) Wins.
Plotted for each statistical metric are the distribution of career totals for players whose career ended in
the periods 1920-1960 and 1960-2000.
 The pairs of distributions are all qualitatively similar, with the exponential cutoffs occurring at the same critical value, indicating that the competition level in baseball has been relatively constant with respect to these career metrics.
   }
 \label{4g}
\end{figure}
In Fig. \ref{4g} we plot the distribution of career batting and pitching
totals for all players who ended their careers between the years of  1920-1960 and 1960-2000 (we restrict our analysis to completed careers). 
Separating players into two subsets allows us to compare careers belonging to each era, where 1961 marks the beginning of the first expansion era in major league baseball.
We also find truncated power-law behavior with exponent $\alpha
\approx 1$ for all major career metrics.
This should not be too surprising since each opportunity (defined in
this paper as an {\it at-bat} or {\it out-pitched}) is capitalized upon
at a player's personal rate (defined in this paper as his {\it
prowess}); each success then contributes to the player's career statistical tally. 
Thus, the exponent from the career longevity power-law should carry over naturally into the density functions of career metrics \cite{BB2}. 
In the case of batting statistics, we make no
distinction between pitchers and other fielders who are on record for
their at-bats.  One can also do a statistical analysis on players who do
not arise in the pitching database, but the distributions are not
qualitatively different.  Thus, career longevity measured in at-bats
indicates that there is a large disparity between the ``iron-horses" and
the ``one-hit wonders".  It is perplexing that there is such a wide range of career lengths
despite the typical prowess that distinguishes the upper echelon of
baseball talent. It should also be noted that in the game of
baseball there are two classes of pitchers, those that start games, and
those that finish games.  Pitchers of the first type have routine
schedules, pitching once every four or five games in a maintained
rotation.  Pitchers of the second type pitch more frequently, with game sessions that are shorter, 
hardly ever exceeding 2 innings (6 outs pitched).  
Despite these two classes of pitchers, the longevity measure
of outs-pitched does not have any evidence of bimodal behavior.  One can
even notice the fluctuations in the beginning of the distribution for
outs-pitched with sharp peaks corresponding to 1 inning (3 outs) and 2
inning (6 outs) stints.  Comparing pitchers and batters, there is the
remarkable similarity in power-law exponent corresponding to longevity,
following from the fact that it is very difficult to reach, and to
remain, at the major league level.  Moreover, the distributions are
nearly equivalent, with the exponential cutoff occurring at
approximately the same value.  This justifies both the 3000-hit and the
3000-strikeout benchmarks for both batters and pitchers, and suggest
that career longevity results from a universal mechanism that is
invariant with respect to player type.
Baseball relies on precision play, requiring quick
physical and mental reflex.  The flow of the game is characterized by
periods of lull, interlaced with bursts of activity, with both offense
and defense capitalizing on sprinting, diving, and sliding plays \cite{Outfielders}.
In addition, throwing a baseball is very strenuous on the arm.  Thus,
although not a contact sport in the sense of hockey, rugby, or American
football, baseball players are still prone to injury, some of which are
career-ending.  The perpetual hazard of career-ending replacement or injury provides
the main ingredient for explaining the observed power laws.  
In Ref.\cite{BB2} we propose a simple stochastic mechanism for career longevity
in professional sports which reproduces both the power-law behavior and the exponential cutoff.
We follow the explanation of Reed {\it et al.} \cite{ReedH}, which shows that
stochastic processes with exponential growth produce power-laws when the
process is subject to random stopping times.\\
 \begin{figure}
 \centering
\includegraphics[width=0.45\textwidth]{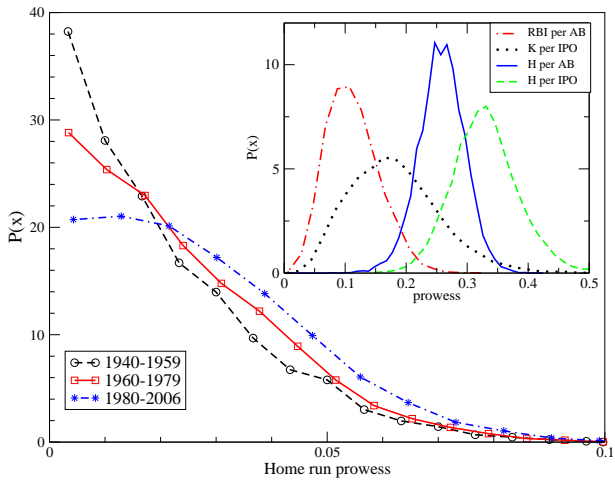}
  \caption{
 Probability density function of seasonal home-run prowess (pitchers
are excluded from this analysis).
 The exponential distribution, representing players in the years 1920-1979, is
skewed towards smaller rates, indicating that even at the major league level, the
ability to consistently hit home runs is rare. Players from the last 25 years have increased their ability to hit home runs, possibly as a result of modern training regimens, performance-enhancing drugs, expansion-based dilution of talent, and other hypothetical factors.
(Inset) Probability density function of seasonal prowess for several key metrics over the seasons 1920-2000.
These are centered distributions with a mean $\langle x \rangle$, and standard deviation $\sigma$, that define the talent level in the major leagues. }
\label{prowess}
\end{figure}
\begin{figure}
\centering
\includegraphics[width=0.45\textwidth]{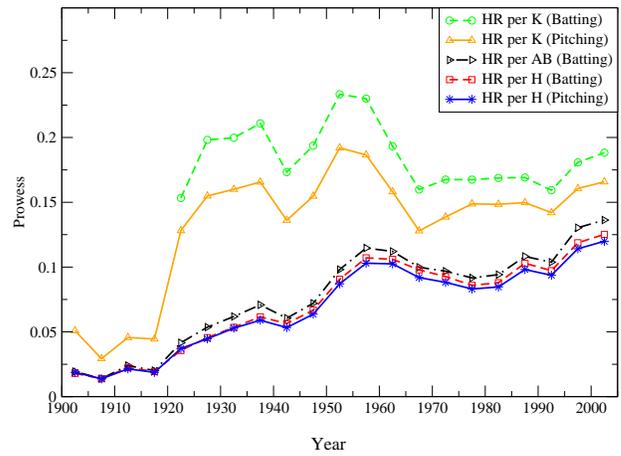}
\caption{Average league prowess $\langle P \rangle$ calculated for 5-year periods starting in 1900. 
The trendline for HR/AB has been multiplied by a factor of 4 for clarity. 
The difference in the trendlines for Hr per Hit for batter and pitcher are essentially constant over the years.
Following from the form of our weighted average, this suggests that the changes in home run prowess are not due to fluctuations in the talent pool arising from expansion dilution. 
If more home runs were being hit by veteran batters against poor pitching, then a spread would appear, assuming that poor pitchers don't stay at the major league level very long.
Instead, we see that veteran hitters (with a large fraction of the total hits) are hitting more home runs off of veteran pitchers (with a large fraction of hits surrendered, and thus many innings pitched).
}
\label{yearlyprowess}
\end{figure}
In Fig. \ref{prowess} we analyze seasonal home run prowess, defined as the
rate per at-bat in which a particular player hits a home run.  Pitchers
are excluded from this analysis.  We also restrict our analysis to
players who exceed $N$ appearances in a given season, and use $N=100$ to
eliminate statistical fluctuations arising from short-lived success.
The seasonal prowess distributions for some common batting and pitching
metrics are relatively unskewed, defined by a characteristic
standard deviation around a central mean (Fig. \ref{prowess}, inset).
Thus, there is a typical success rate that defines not
only the players, but also the {\it relative} level of competition between pitcher and batter at the major league
level.  In contrast, the seasonal prowess distributions for home runs
are more exponentially distributed (Fig. \ref{prowess}).  These distributions are skewed
towards small values, indicating that it is rare for players to have
prowess that consistently produces home runs.  We also note that the
distributions for home-run prowess over the past 26 years reveals a
shift towards players with higher home-run ability.  This increase in home-run prowess could result from
modern {\it natural} weight-training programs with or without the use of performance-enhancing drugs.  
Other theories suggest that maple bats, a reduced strike-zone, and league expansion all contribute to the increased home run performance of modern players in the ``Steroid Era".
A recent study by R. Tobin \cite{SteroidsTobin} demonstrates that a reasonable increase in a player's muscle-mass, say a 10 percent increase, can produce a significant increase in home run production, ranging from 30-70 percent increase, depending on systemic parameters. 
Thus, our findings are consistent not only with the factual revelations of the Mitchell Report, but also with the aforementioned Monte-Carlo simulations.

It has been known for some time that home run rates over the last two decades have been rising \cite{SportsEco}, accompanied by home run records falling.
In Fig. \ref{yearlyprowess} we plot the average prowess of several metrics over 5 year windows from 1900-2005 in order to investigate the evolution of home run prowess.
If in a single season player $i$ has prowess $P_{i} = x_{i}/y_{i}$ \ , then we compute the weighted average over all players
\begin{equation}
 \langle P \rangle_{T} = \frac{\sum_{i} x_{i}}{\sum_{i} y_{i}} = \sum_{i \in T} w_{i}P_{i} \nonumber 
 \end{equation}
 where
 \begin{equation}
  w_{i} = \frac{y_{i}}{\sum_{i}y_{i}} \nonumber
  \end{equation}
The index $i$ runs over all individual player seasons during the period $T$, and $\sum_{i}y_{i}$ is the total number of events $y$ during the same period.
The first era of increasing home-run prowess followed the 1920 revision
of the rules (such as the outlaw of the ``spit-ball") which made the
batter and pitcher more equally competitive.  This was followed by the
emergence of sluggers such as Babe Ruth, who popularized the herculean
feat of hitting home runs \cite{KB99}.
The first expansion era 1961-1969 saw 8 new teams, accompanied by a decrease in average home-run prowess.
It is important to note that expansion within a league has two main effects. On the player level, expansion dilutes the talent in pitching and batting. 
This allows excellent players to take advantage of their weaker foe, and has been proposed as a possible factor responsible for the increased home run rate during the 1990's \cite{NYT}.
On the team level, the authors of  Ref.~\cite{parity} show that the level of team competition measured in team-versus-team upset probability increases during expansion eras.
The second expansion era 1993-1998 saw 4 new teams, accompanied by an increase in average home-run prowess following approximately 20 years of stagnancy.
\begin{figure}
  \centering
\includegraphics[width=0.4\textwidth]{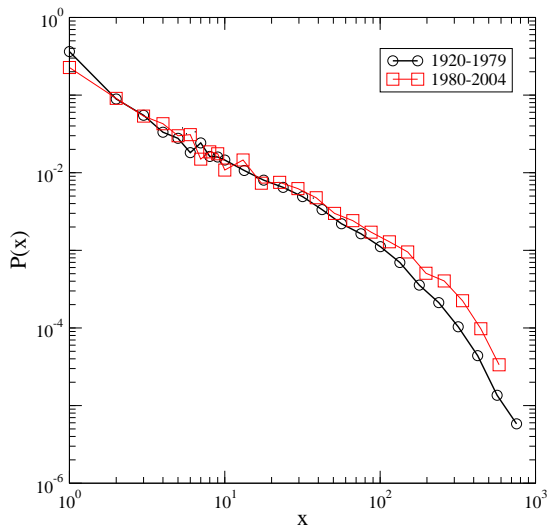}
  \caption{Statistical evidence in career home run distributions consistent
with performance enhancement drugs. Probability density function of home runs hit over a player's career
ending in two different time periods, before and after 1980 (pitchers are
excluded). More home runs are being collected in the extreme part of the distribution by
individuals ending their careers in the last 25 years, 1980-2004, marked by the
``steroids era".
  }
  \label{hr}
\end{figure}

Because career statistics serve as key metrics for
classifying players and establishing their historical legacy, we separate the players in Fig. \ref{hr} into two subsets, players ending their careers before and after 1980, in order to compare career home run totals.
We find that the last 25 years
account for many more players with large career home-run tallies.
Interestingly, there is similar evidence in the strikeout tallies of
pitchers (Fig. \ref{4g}c), which suggests that modern sluggers may be
``swinging for the fences" with reckless abandon.  
We should note that the difference in career statistics for strikeouts is confined to the highly sensitive exponential tail, whereas the differences in the career statistics for home runs extends into the bulk of the distribution.
The use of steroids
was most recently documented in the Mitchell Report \cite{mitchell}, a
two-year investigation into the use of performance-enhancing drugs in
major league baseball.  This paper reports a {\it lower-limit} to the
extent of steroid use in major league baseball at 5 percent, the results
of a set of anonymous 2003 blood test that confirmed the widespread use
of performance-enhancing drugs among major league players.  Other
Mitchell Report assessments, based on personal accounts, suggest much
higher percentages of steroids use in professional baseball.  Steroids
and other performance-enhancing drugs can be used for two general
reasons, to gain strength and to reduce recovery-time from both workouts and injury.  One might
expect that performance-enhancing drugs would raise the level of
competition across the board, for pitchers as well as batters, since
both increased strength and speedy recovery can contribute to high
career tallies. However, in our analysis of career statistics, we see evidence for a competitive advantage
mainly in the case of home runs. This suggests that the level of
competition between pitcher and batter is tipping in the favor of the
batter.\\

Major league baseball is a unique sport, relying on team play while also
maintaining a platform for individual play, namely pitcher versus
batter.  It also has a deep developmental minor league system that
filters out the best talent, and serves as a emergency source for
randomly depleted team rosters.  This provides an explanation for the
abundance of hitters and pitchers who experience both their debut and
finale in the same game. In \cite{BB2} we analyze career longevity in
Korean baseball, American basketball, and English football.  
We find the same power-law behavior with exponential cutoff governing career statistics in all these professional sports, and provide a stochastic mechanism to explain the distribution of career length in competitive environments that are subject to random exit times.
It should also be noted that performance-enhancing drugs are not a
problem unique to American baseball.  A separate study of English
football also revealed widespread use of performance-enhancing drugs
\cite{BritFB}.  Moreover, it is not just a problem pertaining to professionals, but amateurs and adolescents as well \cite{Ped}, as
 performance-enhancing drugs are the core of a pandemic
that not only poses personal health risk, but also places the integrity
of sports in jeopardy \cite{Surg,NEJM}.
And finally, crossing over into the academic world, a recent study in the journal {\it Nature}\cite{CED1} reveals that  cognitive-enhancing drugs are prevalent in the sciences, 
and pose the same ethical questions that apply to accomplishments in sports \cite{CED2}.

\end{document}